\documentclass[aps,prb,twocolumn,showpacs]{revtex4}

\usepackage{amsmath}
\usepackage{amssymb}
\usepackage{graphicx}

\newcommand*{\cZ}{{\cal Z}}
\newcommand*{\cC}{{\cal C}}
\newcommand*{\cA}{{\cal A}}
\newcommand*{\cD}{{\cal D}}
\newcommand*{\cH}{{\cal H}}
\newcommand*{\cG}{{\cal G}}
\newcommand*{\cT}{{\cal T}}
\newcommand*{\cP}{{\cal P}}

\begin{document}


\title{Localization from $\sigma$-model geodesics}


\author{A.~Lamacraft}
\affiliation{Department of Physics, Princeton University, Princeton, NJ
  08544, USA}
\email[]{alamacra@princeton.edu}

 \author{B.\,D.~Simons}
 \affiliation{Cavendish Laboratory, Madingley Road, Cambridge CB3\ OHE, UK}

 \author{M.\,R.~Zirnbauer}
\affiliation{Institut f\"ur Theoretische Physik, Universit\"at zu
K\"oln, K\"oln, Germany}

\date{February 24, 2004}

\begin{abstract}
We use a novel method based on the semi-classical analysis of
$\sigma$-models to describe the phenomenon of strong localization
in quasi one-dimensional conductors, obtaining the density
of transmission eigenvalues. For several symmetry
classes, describing random superconducting and chiral
Hamiltonians, the target space of the appropriate $\sigma$-model
is a (super)group manifold. In these cases our approach turns out
to be exact. The results offer a novel perspective on
localization.
\end{abstract}


\maketitle


\section{Introduction}

Beginning with the early work of
Thouless~\cite{edwards_thouless,thouless}, quasi one-dimensional
conductors have provided a valuable arena in which to explore the
influence of quantum interference effects on the transport
properties of weakly disordered phase coherent conductors. By the
early 80's, a complete scaling theory of localization in
multi-mode wires had been formulated by Dorokhov~\cite{dorokhov}
(and developed later by Mello, Pereyra and Kumar~\cite{mpk}) as a
Brownian motion of eigenvalues of the transmission matrix --- the
`DMPK equation' (for a recent review see, e.g.,
Ref.~\onlinecite{beenakker}). For several symmetry classes,
analytic  results for the low moments of conductance were obtained
in both the metallic and strongly localized
regimes~\cite{beenakker,bfgm,macedo2}. Lately, a third, and potentially
more versatile approach has been developed to investigate quantum
transport in (multi-terminal) disordered conductors. By formally
relating the transmission eigenvalue distribution to a
`multi-component Green's function' (see below), Nazarov has
shown~\cite{nazarov} that known results in the metallic limit can
be inferred from the equations of motion of the average
quasi-classical Green function~\cite{eilenberger,lo,usadel}.

The transport properties of a conductor are fully specified by the
transmission matrix ${\bf t}$~\cite{buttiker}. In particular, the
Landauer-B\"uttiker formula allows the conductance to be expressed
through eigenvalues $\cT_n$ of the `squared' transmission matrix,
\begin{eqnarray*}
  G=\frac{e^2}{h}\mathrm{tr}\,{\bf t}^\dagger {\bf t}=\frac{e^2}{
  h} \sum_n \cT_n\,.
\end{eqnarray*}
The statistics of the $\cT_n$ follow the universal Dorokhov
distribution, valid in the metallic limit (i.e. where the dimensionless conductance 
$\mathfrak{g}\equiv hG/e^2\gg 1$):
\begin{equation}
\rho(\cT\gtrsim e^{-2L/\ell})=\Big\langle \sum_n \delta(\cT-\cT_n)
\Big\rangle=\frac{\mathfrak{g}}{2}\frac{1}{\cT\sqrt{1-\cT}}\,.
\label{dorokhov}
\end{equation}
By solving the corresponding quasi-classical equation in the
diffusive limit, Eq.~(\ref{dorokhov}) has been established under
very general conditions. Yet, while the method introduced by
Nazarov is capable of describing accurately the transmission
eigenvalue distribution in the metallic phase, the method fails to
account for the phenomena of weak and strong localization.

In recent years, the development of a field theoretic technique to
explore phase coherence phenomena in disordered conductors has
exposed the  strengths and limitations of the quasi-classical
scheme. Weakly disordered conductors and superconductors can be
classified into one of ten symmetry classes~\cite{suprev}, their
spectral and transport properties specified by a field theory of
nonlinear $\sigma$-model type~\cite{efetov,az}. It was realized
that within the formally exact framework provided by the field
theory, the quasi-classical equations of motion represent a
saddle-point or mean-field approximation, involving only a single
field configuration~\cite{mk,ast}. Indeed, drawing on the
substantial literature on the quasi-classical equations in the
superconducting context (the Eilenberger~\cite{eilenberger} and
Usadel~\cite{usadel} equations), this identification simplified
substantially the analysis of physical realizations of the `novel
symmetry' classes of disordered superconductors introduced by
Alt\-land and Zirnbauer~\cite{az}. The absence of localization in
the quasi-classical theory follows then from the neglect of
fluctuations about this field configuration.

Describing localization within the $\sigma$-model is however far
from trivial. While weak localization corrections to transport can
be developed as a systematic perturbation theory, the transition
to strong localization is signalled by the growth of contributions
non-perturbative in the conductance. Indeed, a calculation by
Rejaei~\cite{rejaei} of the exact transmission eigenvalue
distribution relied on a mapping of the $\sigma$-model for the
quasi one-dimensional system onto a heat kernel (see also 
Ref.~\onlinecite{macedo1}). Such an analysis affords little intuition into the origin of localization within the framework of the $\sigma$-model. In particular, can the
transition (or, in quasi one-dimension, the crossover) to strong
localization be understood in terms of certain $\sigma$-model
field configurations?

A preliminary answer to this question was provided in an
insightful work by Muzykantskii and Khmel'nitskii~\cite{mk}. Using
the field theoretic approach, it was demonstrated that the
long-time response of a weakly disordered quasi one-dimensional
wire to a voltage step was controlled by spatially inhomogeneous
saddle-point field configurations of the $\sigma$-model action.
Non-perturbative in the conductance, these field configurations
were associated with rare ``anomalously localized'' states
embedded deep within the metallic phase. Whether these states
provided a caricature of states near the Anderson transition
remains to date the subject of debate. Later, in a related
activity, it was shown the nucleation of localized quasi-particle
states inside the gap of a weakly disordered symmetry broken
superconductor were ascribed to spatially inhomogeneous
saddle-point configurations of the action~\cite{ls}.

In this paper we will show that, for a range of symmetry classes,
the physics of strong localization is captured \emph{exactly} by
the inclusion of a set of new saddle points combined with
their associated Gaussian fluctuations. This recalls a previous
investigation by Andreev and Altshuler~\cite{aa} of energy level
correlations in symmetry broken (zero-dimensional) chaotic systems
--- random matrix ensembles belonging to the unitary symmetry
class. There it was shown that the two-point correlator of the
density of states could be fully recovered from the inclusion of
two saddle points together with their associated Gaussian
fluctuations. In both cases, the coincidence can be traced to the
property of semiclassical exactness shared by the different
$\sigma$-model field theories. The aim of the present paper is to
elucidate this principle in the quasi one-dimensional system
comparing results for transmission eigenvalue distributions to
those obtained from heat kernel methods. The novel
view of localization presented here may inform the treatment of more
complicated problems.

Following Rejaei, our analysis rests on a relation which allows
the ensemble average of the generating function of the
transmission matrix ${\bf t}{\bf t}^{\dagger}$ (at a particular
energy) of a quasi one-dimensional sample
\begin{equation} \label{tgen}
\cZ(\phi,\theta)=\left\langle\frac{\textrm{det}(1-\sin^2(\theta/2){\bf t}
{\bf t}^{\dagger})}{\textrm{det}(1-\sinh^2(\phi/2){\bf t}{\bf t}^{\dagger})}
\right\rangle\;,
\end{equation}
to be presented as a partition function of a nonlinear
$\sigma$-model of the appropriate symmetry class. The density of
transmission eigenvalues may then be extracted using the relation
\begin{equation}
  F(\phi) \equiv \frac{\partial}{\partial \theta} \cZ(\phi,\theta)
  \big|_{\theta = i\phi} = \sum_n \left\langle \frac{-i\sinh \phi}
  {\cosh \phi_n + \cosh \phi} \right\rangle,
  \label{Fdef}
\end{equation}
where $\cT_n = 1 / \cosh^2(\phi_n / 2)$ denote the eigenvalues of
the matrix ${\bf t}{\bf t}^\dagger$. From this function one can
infer the transmission eigenvalue density through the relation
\begin{equation}\label{density}
    \rho(\phi) \equiv \sum_n \big\langle \delta(\phi - \phi_n)
    \big\rangle = \frac{1}{2\pi} \big( F(\phi +
    i\pi) - F(\phi - i\pi) \big) \;.
\end{equation}
The moments of the transmission matrix can be cast in terms of the
retarded and advanced Green's functions $G^{R/A}\equiv(\epsilon\pm
i\delta-H)^{-1}$ of the microscopic Hamiltonian of the wire as
\[\mathrm{tr}({\bf t}{\bf t}^{\dagger})^n=\mathrm{tr}_{{\bf r}}(\hat v_L
G_{\epsilon}^A \hat v_R G_{\epsilon}^R)^n\;,\]
where $\hat v_{L/R}$ denotes the current operator through left and
right cross-sections at $x=0$ and $x=L$ respectively, and the
trace on the right hand side runs over spatial coordinates.
Following Nazarov, it is helpful to note that the determinants in
Eq.~(\ref{tgen}) can be recast in terms of a multi-component
Green's function according to the relation
\begin{eqnarray*}
\textrm{det}(1-\gamma_1\gamma_2 {\bf t}{\bf t}^{\dagger})&=&\textrm{det}
(1-\gamma_1\gamma_2\hat v_L G_{\epsilon}^A \hat v_R G_{\epsilon}^R)\\*
&=&\textrm{det}\begin{pmatrix}1 & \gamma_1 G_{\epsilon}^R\hat v_L \cr
\gamma_2 G_{\epsilon}^A \hat v_R & 1 \end{pmatrix}\\*
&\propto&\textrm{det}\begin{pmatrix}\epsilon-H+i\delta& \gamma_1 \hat v_L
\cr \gamma_2 \hat v_R & \epsilon-H-i\delta \end{pmatrix},
\end{eqnarray*}
where, for example, in the case of the numerator of (\ref{tgen}),
$\gamma_1 \gamma_2\equiv\sin^2(\theta/2)$ . Thus the generating
function $\cZ(\phi,\theta)$ may be related to the Green's function
of an enlarged Hamiltonian, which includes an off-diagonal `vector
potential'.

With this representation in hand, the outline of the remainder of
the paper is as follows. In the next section we present a complete
analysis of the localization properties of Class $C$I, one of the
superconducting symmetry classes introduced by Altland and
Zirnbauer~\cite{az}. We outline the $\sigma$-model description of
this symmetry class and reproduce Nazarov's calculation in this
context, before introducing the full set of saddle-point
trajectories - geodesics on the target space - and performing a
complete semiclassical calculation. From the latter we will
extract the transmission eigenvalue density and mean conductance.
The exactness of the results obtained are then verified by heat
kernel methods. In the following section, the techniques developed
above will be applied to two additional symmetry classes that are
amenable to such a treatment. Finally, in the last section, we
will draw conclusions.

\section{Class $C$I} \label{sec:semiclCI}

Superconductors which exhibit both time reversal symmetry and spin
rotation invariance belong to symmetry class $C$I. By introducing
the familiar Nambu doublet of electron and hole operators in what
we will refer to as the `particle-hole' (ph) space, in the
mean-field approximation, the quasi-particle Gor'kov or
Bogoliubov-de Gennes Hamiltonian $\hat\cH_{\mathrm{mf}}$ can be
written as
\begin{eqnarray*}
\hat\cH_{\mathrm{mf}}=\frac{1}{2}\sum_{ij}
\begin{pmatrix}
\hat c^{\dagger}_{i\uparrow} & \hat c^{\vphantom{\dagger}}_{i\downarrow}
\end{pmatrix}
\overbrace{
\begin{pmatrix}
h_{ij} & \Delta_{ij} \cr \Delta_{ij} & -h_{ij}
\end{pmatrix}}^{H_{ij}}
\begin{pmatrix}
\hat c^{\vphantom{dagger}}_{j\uparrow} \cr \hat c^{\dagger}_{j\downarrow}
\end{pmatrix},
\end{eqnarray*}
where $h_{ij}$ and $\Delta_{ij}$ represent real symmetric matrix
elements corresponding to the non-interacting single-particle
Hamiltonian of the quasi one-dimensional disordered wire and the
superconducting order parameter respectively. With this structure,
the matrix Hamiltonian $H$ satisfies the symmetry relations
\begin{equation} \label{phsym}
 H=H^T=-\cC H^T \cC^{-1}\qquad \cC=\begin{pmatrix} 0 & 1 \cr -1 & 0
\end{pmatrix}_{\mathrm{ph}}\;,
\end{equation}
which in turn implies the identity
\[ G^R_{\epsilon}=-\cC(G^A_{-\epsilon})^T\cC^{-1}\;, \]
between the retarded and advanced Green's functions.

\subsection{Partition function}

With this definition, the ratio of determinants which specifies
the generating function~(\ref{tgen}) can be conveniently expressed
via a supersymmetric field integral, with the numerator arising
from a fermionic integral and the denominator from its bosonic
counterpart. Since these are not identical, the corresponding
partition function will inherit boundary conditions which break
the supersymmetry. To identify the soft diffusion modes of the
superconducting system, it is convenient to effect a doubling of
the field space of the supervectors $\psi$ to include a
`charge-conjugation' (cc) space~\cite{zirnsuper}
\begin{equation} \label{ccdoubling}
\Psi=\frac{1}{\sqrt{2}}\begin{pmatrix} \psi \cr \cC\bar\psi^T
\end{pmatrix}_{\mathrm{cc}} \qquad \bar\Psi=\frac{1}{\sqrt{2}}
\left(\bar\psi, -\psi^T\cC^{-1}\right)_{\mathrm{cc}}.
\end{equation}
Focusing on the transmission eigenvalue distribution at
$\epsilon=0$, we notice that if $G^R_0\sim \langle
\psi\bar\psi\rangle$ is the $(1,1)$ component of the matrix
Green's function $\cG$ in the cc space, then the advanced
component $G^A_0\sim -\cC\langle\bar\psi^T\psi^T\rangle\cC^{-1}$
is just the (2,2) component. Thus the enlarged structure required
is just the usual cc space. Altogether, this leads to the
representation
\begin{eqnarray*}
&&\frac{\textrm{det}(1-\gamma_1\gamma_2 {\bf t}{\bf t}^{\dagger})}{\textrm{det}
(1-\zeta_1\zeta_2 {\bf t}{\bf t}^{\dagger})}\nonumber \\&&\qquad
=\int \cD\Psi\cD\bar\Psi\exp\Big[\frac{i}{2}\int d{\bf r}\bar\Psi
\overbrace{[i\delta\Sigma_3-\cH]}^{\cG^{-1}}\Psi\Big]
\end{eqnarray*}
where
\begin{eqnarray} \label{enlargedH}
&&\cH=H\otimes\openone_{\mathrm{bf}}\otimes\openone_{\mathrm{cc}}\\
&&\qquad +\hat v_L
\Gamma_1\otimes\Sigma_+\otimes\openone_{\mathrm{ph}}+\hat v_R\Gamma_2\otimes
\Sigma_-\otimes\openone_{\mathrm{ph}}\;,\nonumber
\end{eqnarray}
where $\Gamma_{1,2} = \mathrm{diag}(\zeta_{1,2}, \gamma_{1,2}
)_{\mathrm{bf}}$, and $\Sigma_i$ denote the Pauli matrices in the
cc space.

Once cast in the form of a functional field integral, it is a
straightforward (if somewhat lengthy) and standard procedure to
show that the low-energy properties of the partition function are
contained with an effective field theory of nonlinear
$\sigma$-model type (for a detailed discussion of the explicit
derivation, we refer to one of the many standard references, e.g.
Ref.~\cite{zirnsuper,ast}). Two points should be remarked upon.
The first is that, since Class $C$I has two symmetries --
particle-hole and time reversal (tr) symmetry (see
Eq.~\ref{phsym}) -- the supervector needs to be doubled
\emph{twice}. The size of the supermatrix field $Q(\mathbf{r})$
that appears at an intermediate stage in the derivation is thus
$16\times 16$ (ph$\times$cc$\times$tr$\times$bf), but this is
quickly reduced to the $8\times 8$ field $q$ by $Q=\sigma_3 q$ on
the saddle-point manifold ($\sigma_i$ denote the Pauli matrices in
ph space), with the ph space then disappearing from view.
Secondly, following Rejaei, the `vector potential' associated with
the coupling of the wire to the external leads, can be absorbed
into a rigid boundary condition on the field integral. Taking the
metallic contacts at $x=0,L$ to be to clean normal conducting
leads, the ensemble averaged partition function assumes the form
(with $\hbar=1$)
\begin{widetext}
\begin{eqnarray}
\left\langle \frac{\textrm{det}(1 - \gamma_1\gamma_2 {\bf
t}_{\epsilon} {\bf t}^{\dagger}_{\epsilon})}
{\textrm{det}(1-\zeta_1\zeta_2 {\bf t}_{\epsilon}{\bf
t}^{\dagger}_{\epsilon})}\right\rangle = \int_{q(0) =
\Sigma_3}^{q(L) = S\Sigma_3 S^{-1}} \cD q \exp\left( \frac{\pi\nu
D}{8} \int d{\bf r}\, \mathrm{STr} \left[\nabla q 
\right]^2 \right)\;, \label{sigma}
\end{eqnarray}
\end{widetext}
where the field integral is over $8\times 8$ supermatrix fields
$q$ subject to the nonlinear constraint $q^2({\bf r})=\openone$.
Finally, the supersymmetry breaking source terms enter the
boundary condition, through the rotation matrix
\begin{displaymath}
S=\exp\left(i\Gamma_2\otimes\Sigma_-\otimes\openone_{\mathrm{tr}}\right)
\exp\left(i\Gamma_1\otimes\Sigma_+\otimes\openone_{\mathrm{tr}}\right) \;.
\end{displaymath}
Specifically, if we choose
\begin{eqnarray*}
&&\gamma_1=\frac{1}{2}\sin \theta\;,\qquad \gamma_2=\tan \theta/2\\
&&\zeta_1=\frac{i}{2}\sinh \phi\;,\qquad \zeta_2=i\tanh \phi/2\;,
\end{eqnarray*}
then the boundary condition at $x=L$ is
$q(L)=\mathrm{diag}(q_{\mathrm{bb}}(L),
q_{\mathrm{ff}}(L))_{\mathrm{bf}}$ where
\begin{eqnarray}\label{bc}
&&q_{\mathrm{bb}}(L)=\begin{pmatrix} \cosh \phi  & \sinh \phi \cr -\sinh \phi
&-\cosh \phi \end{pmatrix}_{\mathrm{cc}}\otimes\openone_{\mathrm{tr}}\;,\nonumber\\
&&q_{\mathrm{ff}}(L)=\begin{pmatrix} \cos \theta & -i\sin \theta \cr i\sin
\theta & -\cos \theta \end{pmatrix}_{\mathrm{cc}}\otimes\openone_{\mathrm{tr}}\;,
\end{eqnarray}
and we have obtained the required form of Rejaei's relation (\ref{tgen}). An
alternative approach that matches Nazarov's orginal formulation is to use the
explicit expression for $F(\phi)$ that follows from Eq.~(\ref{Fdef}) and
relate it directly to the Green's function in the cc space, i.e.
\begin{eqnarray}
F(\phi)&=&-\frac{i}{2}\sinh \phi\left\langle\mathrm{tr}\left[{\bf t}
{\bf t}^{\dagger}\left( 1+\sinh^2(\phi/2){\bf t}{\bf t}^{\dagger}
\right)^{-1}  \right]\right\rangle \nonumber \\*
&=&\left\langle\mathrm{tr}_{\bf r}\left[ \hat v_R  \cG^{12}_{\mathrm{bb}}
\right]\right\rangle\big|_{\theta=i\phi}\nonumber\\*
&=&-\frac{i\pi\nu D}{2} \cA\left\langle \mathrm{tr}_{\mathrm{tr}}
\left(q\partial_x q\right)^{12}_{bb}\big|_{x=0}
\right\rangle_q\big|_{\theta=i\phi},
\label{nazF}
\end{eqnarray}
where $\langle\cdots\rangle_q$ denotes an average with respect to
the $\sigma$-model action~(\ref{sigma}), $\cA$ is the
cross-sectional area, and $\nu$ is the single particle density of
states. $\mathrm{tr}_{\mathrm{tr}}$ denotes the trace over the tr
space. Henceforth, we will work only with the one-dimensional
field $q(x)$, assuming that the width of the wire is much smaller
than the localization length.

\subsection{Nazarov's calculation}

To develop a theory of localization from the field integral, it is
first useful to establish contact with the theoretical framework
introduced by Nazarov. As discussed in the introduction, the
quasi-classical theory formulated by Nazarov is contained within
the saddle-point structure of the present action: Specifically, a
variation of the action with respect to $q$ subject to the
nonlinear constraint, $q^2=\openone$, obtains the saddle-point
equation
\[\partial_x(q\partial_x q)=0,\]
which expresses conservation of the matrix current $q\partial_x
q$. Indeed, associating $q$ with the average quasi-classical Green
function, the saddle-point equation can be interpreted as the
generalization of the quasi-classical equation derived in
Nazarov's early work to the superconducting wire.

The simplest solution satisfying the given boundary conditions is
block diagonal in the bf space and takes the form
\begin{eqnarray}
&&q_{\mathrm{bb}}(x)=\begin{pmatrix} \cosh \frac{\phi x}{L}  & \sinh
\frac{\phi x}{L} \cr -\sinh \frac{\phi x}{L}&-\cosh \frac{\phi x}{L}
\end{pmatrix}_{\mathrm{cc}}\otimes\openone_{\mathrm{tr}},\nonumber\\
&&q_{\mathrm{ff}}(x)=\begin{pmatrix} \cos\frac{\theta x}{L} & -i\sin
\frac{\theta x}{L} \cr i\sin \frac{\theta x}{L} & -\cos \frac{\theta x}{L}
\end{pmatrix}_{\mathrm{cc}}\otimes\openone_{\mathrm{tr}}\;,
\label{nazsol}
\end{eqnarray}
If one inserts this solution into Eq.~(\ref{nazF}), and neglects
field fluctuations around the saddle point, one finds
\[F(\phi)=-i\pi\nu D\cA\phi/L.\]
Then, making use of Eq.~(\ref{density}), one can straightforwardly
recover the Dorokhov distribution (\ref{dorokhov}) for a wire of
dimensionless conductance $\mathfrak{g}=2\pi\nu DA/L$.

However, when derived from the supersymmetric field theory, one
can see that the saddle-point equation presents not just one
solution, but a whole family of solutions in the compact fermionic
sector
\begin{eqnarray}
q^{(n)}_{\mathrm{ff}}(x)=\begin{pmatrix} \cos\frac{(\theta+2\pi n) x}{L} &
-i\sin \frac{(\theta+2\pi n) x}{L} \cr i\sin \frac{(\theta+2\pi n) x}{L} &
-\cos \frac{(\theta+2\pi n) x}{L} \end{pmatrix}_{\mathrm{cc}}\otimes
\openone_{\mathrm{tr}},\nonumber\\* n\in\mathbb Z\,.
\label{windings}
\end{eqnarray}
Thus we see that even at the level of a purely
\emph{semi-classical} analysis, Nazarov's treatment is missing two
elements; corrections associated with fluctuations around the
conventional saddle-point solution~(\ref{nazsol}), and
solutions~(\ref{windings}) that loop multiply around the compact
fermionic sector. We will show that these two `channels' of
corrections are, respectively, responsible for the phenomenon of
weak localization --- as, indeed, one might have guessed from the
connection between the role of fluctuations in the $\sigma$-model
and diagrammatic perturbation theory --- and strong localization.

To uncover this relation, in the following we will undertake the
semi-classical analysis taking into account fluctuations of the
matrix fields at quadratic order around both the conventional and
the non-trivial saddle-point field configurations.

\subsection{Semi-classical calculation}

To perform the field integration over the fluctuations, it is
necessary to review the structure of the target manifold.  For the
minimal ($n = 1$) nonlinear $\sigma$-model associated with the
symmetry class $C$I, the target space --- spanned by matrices $q =
w \Sigma_3 w^{-1}$ --- turns out~\cite{suprev} to be in one-to-one
correspondence with $G = \widetilde{\mathrm{OSp}}(2|2)$, by which
we mean the Lie group $\mathbb{R}_+ \times \mathrm{SU}(2)$
extended to an orthosymplectic Lie supergroup. This group
structure of the target manifold is the reason for the exactness
of the semi-classical approximation~\cite{picken,szabo}.

The correspondence between $q$ and $g \in G$ is described in the
Appendix. The space of matrices~(\ref{bc}) parameterized by $\phi$
and $\theta$ corresponds to the maximal Abelian subgroup $A\in G$
\[  a = \mathrm{diag}(e^\phi, e^{ -\phi}, e^{i\theta},
    e^{-i\theta}) \in A\;,\]
%
while, in terms of group elements $g\in G$, the generating function is
\begin{equation}\label{g_action}
  \cZ(\phi,\theta) = \int\limits_{g(0)={\bf 1}}^{g(T) = a(\phi,\theta)
  }{\cal D}g \, \exp \left( -\frac{1}{8}\int\limits_0^T \mathrm{STr}
  \left[g^{-1}\dot g\right]^2 \, dt \right)\,.
\end{equation}
Here we have switched to the dimensionless variables $t\equiv
x/\xi$, $T\equiv L/\xi$, where $\xi\equiv2\pi\nu D\cA$. (Note that,
with this definition, $T=1/\mathfrak{g}$.) With this parametrization, the
set of saddle-point solutions discussed in the previous section
corresponds to the geodesic trajectories of free particle motion
\begin{equation}
  a_t^{(n)}(\phi,\theta) = \mathrm{diag}(e^{\phi t/T},
  e^{ -{\phi t/T}}, e^{i\theta_nt/T}, e^{-i\theta_nt/T}), \label{spset}
\end{equation}
where $\theta_n=\theta+2\pi n$ with $n\in \mathbb{Z}$.
These configurations are associated with the classical action
\[S_{\mathrm{cl}}^{(n)}=\frac{1}{8}\int\limits_0^T \mathrm{STr}
[(a_t^{(n)})^{-1}\dot a_t^{(n)}]^2 \, dt= (\phi^2 + \theta_n^2) / 4T \]

To evaluate the contribution from Gaussian fluctuations, it is
helpful to set $g_t = a^{(n)}_t \tilde g_t$, and express the
element $\tilde g_t$ through the exponential parametrization
$\tilde g_t\equiv\exp X_t$, with boundary conditions $X_0 = 0
=X_T$. Matrices $X$ satisfying the $\mathrm{osp}(2|2)$ Lie algebra
condition $X = -\varepsilon X^{\rm T} \varepsilon^{-1}$ with
$\varepsilon = {\rm diag}( \sigma_x , i\sigma_y )_{\mathrm{bf}}$
may be written as
\begin{equation}
 X = \begin{pmatrix} d &0 &\alpha &\beta \\ 0 &- d &\gamma &\delta \\
 \delta &\beta &e &b \\ -\gamma &-\alpha &c &-e \end{pmatrix}.
 \label{Xparam}
\end{equation}
Then, to get the desired target space with the non-compact
boson-boson sector $M_{\rm bb} = {\mathbb R}_+$, and compact
fermion-fermion sector $M_{\mathrm{ff}} = \mathrm{SU}(2)$, one
must take the variable $d$ to be real, the variable $e$ to be
imaginary, and the complex variables $b$ and $c$ to be related by
the condition $c = - \bar b$. The functional integration measure
is trivial, ${\cal D}g = {\cal D}\tilde g = {\cal D}X + \ldots$,
up to corrections (of order $X^3$) that will not affect the
present calculation.

Using the parametrization for $X$ given in Eq.~(\ref{Xparam}) one
obtains the quadratic action
\begin{eqnarray*}
  S_{\mathrm q} &=& \frac{1}{8} \int dt \; \mathrm{STr} \left(
   \dot X^2 + \dot X [ (a_t^{(n)})^{-1}
    \dot a_t^{(n)} , X ] \right) \\* &=& \frac{1}{4}\int dt\;
    \left( \dot d^2 - \dot e^2 - \dot b \dot c -
    \frac{i\theta_n}{T} \left( b \dot c - \dot b c \right) \right.
    \\ &&\left. \qquad\qquad + 2 \dot\alpha\dot\delta + \frac{\phi -
    i\theta_n}{T} ( \alpha \dot\delta - \dot\alpha \delta) \right.
    \\ &&\left. \qquad\qquad +
    2 \dot\gamma \dot\beta - \frac{\phi + i\theta_n}{T} ( \gamma
    \dot\beta - \dot\gamma \beta) \right)\;.
\end{eqnarray*}
Performing the Gaussian functional integral over $d$ and $e$ is
the same as computing the kernel for free particle motion on the
two-dimensional Euclidean space ${\rm Lie}\,A$, and the result is
${\rm Det}^{-1}(-\partial_t^2)$. Performing the remaining Gaussian
functional integrals gives determinants in the denominator
(bosons: $b, c = - \bar b$) and numerator (fermions: $\alpha,
\delta$ and $\gamma, \beta$). Collecting all the determinants, one
finally obtains the total fluctuation contribution to the
partition function,
\begin{displaymath}
  \frac{\mathrm{Det}\left( -\partial_t^2 - \frac{\phi + i\theta_n}{T}
  \partial_t \right) \mathrm{Det} \left( -\partial_t^2 + \frac{\phi -
  i\theta_n}{T} \partial_t \right) } { \mathrm{Det} \left( -
  \partial_t^2 \right) \mathrm{Det}\left( -\partial_t^2 + \frac{2
  i\theta_n}{T}\partial_t \right) } \;.
\end{displaymath}

The determinants are to be evaluated on the Hilbert space of
square-integrable functions $L^2([0,T])$ with Dirichlet boundary
conditions.  If $z$ is some complex number, the operator $D_z =
-\partial_t^2 + 2 (z/T) \partial_t $ on that Hilbert space has
eigenfunctions $\sin(k\pi t/T) \, e^{z t / T}$ ($k \in
{\mathbb N}$), with eigenvalues $\left( (k\pi)^2 + z^2 \right) /
T^2$, so its (unregularized) determinant is given by
\begin{displaymath}
  {\rm Det} D_z = \prod_{k = 1}^\infty \left( (k\pi)^2 + z^2
  \right) / T^2 \;.
\end{displaymath}
Taking the logarithmic differential (to kill the $z$--independent
infinity) one obtains
\begin{eqnarray*}
    \delta \ln {\rm Det} D_z &=& \sum_{k = 1}^\infty \frac{2 z
    \delta z}{(k\pi)^2 + z^2} \\ &=& \left( - \frac{1}{z} +
    \sum_{k \in {\mathbb Z}} \frac{z}{(k\pi)^2 + z^2} \right)
    \delta z \\ &=& \left( - z^{-1} + \coth z \right) \delta z
    = \delta ( \ln \sinh z - \ln z ) \;.
\end{eqnarray*}
Therefore,
\begin{displaymath}
  {\rm Det} D_z = \frac{\sinh z}{z} \times {\rm Det} D_0 \;,
\end{displaymath}
and the above ratio of four determinants gives
\begin{displaymath}
   \frac{\sinh \left(\frac{1}{2}(\phi + i\theta_n) \right)}{\frac{1}{2}
   (\phi + i\theta_n)} \, \frac{\sinh \left( \frac{1}{2}(\phi - i\theta_n)
    \right)}{\frac{1}{2} (\phi - i\theta_n)} \left( \frac{\sin \theta_n}
    {\theta_n} \right)^{-1} \;.
\end{displaymath}
Finally, when combined with the exponential of the classical
action and summed over $n$, one obtains the following expression
for the partition function:
\begin{eqnarray}\label{CIZ}
  &&\cZ_T(\phi,\theta) = \sum_{n \in {\mathbb Z}} \frac{\sinh \left(
      \frac{1}{2}(\phi + i\theta_n) \right)}{\frac{1}{2} (\phi + {\rm
      i}\theta_n)} \, \frac{\sinh \left( \frac{1}{2}(\phi - i\theta_n)
    \right)}{\frac{1}{2} (\phi - i\theta_n)}\nonumber \\ && \qquad
    \qquad\qquad\qquad\qquad \times \frac{\theta_n}{\sin
    \theta_n} \, e^{-(\phi^2 +\theta_n^2) / 4T} \;.
\end{eqnarray}

Note that the operator $-\partial_t^2 + (2i\theta_n/T)\partial_t$
starts to exhibit negative eigenvalues as soon as $|\theta_n|$
exceeds $\pi$. This means that the Gaussian functional integral
over $b$ and $c = - \bar b$ does not exist in those cases. Thus
what we have done --- evaluating the functional integral in the
Gaussian approximation around all of the saddle points (without
paying attention to question of existence), and summing
contributions --- was a purely formal calculation. Nevertheless,
the answer obtained in this way turns out to be exact! Indeed, the
validity of this expression can be established both directly and
indirectly. In the following, we will motivate this conclusion by
investigating the transmission eigenvalue density and the mean
conductance from the partition function. Then, by interpreting the
field integral as a heat kernel, we will find an alternative exact
method of computation.

\subsection{Eigenvalue density}

With the partition function in hand, it is a straightforward matter to
compute the eigenvalue density making use of the
relations~(\ref{tgen})-(\ref{density}),
\begin{widetext}
\begin{equation}\label{xdensity}
  \rho^{C\mathrm{I}}_{T}(\phi) = \frac{1}{2T} - \frac{1}{2(\phi^2 + \pi^2)}
  - \sum_{n \not= 0} \frac{ e^{- n (n+1)\pi^2 /T} }{2\pi^2 n} \, {\rm Re} \,
  \frac{ \phi + i\pi (2n+1) }{ \phi + i\pi (n+1) } \, e^{
    i \pi n \phi / T} \;.
\end{equation}
\end{widetext}
The first term of this expression is just the Dorokhov
distribution, and originates from the $n=0$ (Nazarov's)
configuration. In this context, the second term simply denotes the
weak localization correction associated with the $n=0$
configuration. All higher-order terms of the perturbation
expansion in powers of $T=1/\mathfrak{g}$ turn out to vanish identically,
a direct manifestation of semi-classical exactness. Indeed, the
remaining terms in the result~(\ref{xdensity}) depend
non-analytically on $T$ as $e^{-1/T}$, and arise from the
non-trivial saddle-point configurations.

\begin{figure*}
\begin{center}
\setlength{\unitlength}{4.25in}
\begin{picture}(1,0.25)(0.2,0)
  \put(0.02,0){\includegraphics[width=0.6\unitlength]{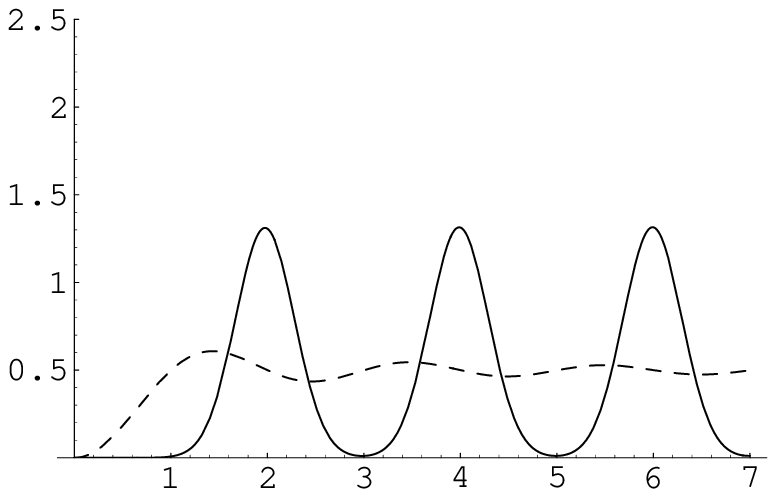}}
  \put(0.38,-0.05){\makebox(0,0)[b]{$\phi/T$}}
  \put(0,0.22){\rotatebox{90}{\makebox(0,0)[t]{$T\;\rho^{CI}_T(\phi)$}}}
  \put(0.67,0){\includegraphics[width=0.6\unitlength]{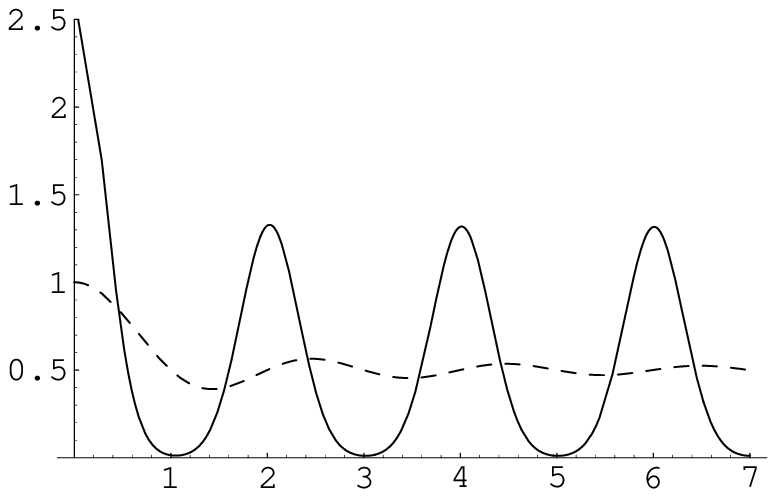}}
  \put(1.03,-0.05){\makebox(0,0)[b]{$\phi/T$}}
  \put(0.65,0.22){\rotatebox{90}{\makebox(0,0)[t]{$T\,\rho^{DIII}_T(\phi)$}}}
\end{picture}
\end{center}
\caption{Transmission eigenvalue densities for the classes $C$I
and $D$III at $T= 0.01$ (weak localization -- dashed line) and $T=
20$ (strong localization -- solid line).  Note that although
eigenvalue `crystallization' occurs in both classes, localization
in class $D$III is absent due to the persistence of an eigenvalue
near $\phi = 0$ (or $\cT = 1$).\label{fig:densities}}
\end{figure*}

A non-trivial check on the validity of the expression follows from
the requirement that $\rho_T(0) = 0$ for all $T > 0$. (One can
infer this condition from the joint probability density given by
Brouwer \emph{et al.}~\cite{bfgm}.) The plot of the density
function $\rho_T^{C{\rm I}}(x)$ in Fig.~\ref{fig:densities} shows
the characteristic `crystallization' of transmission eigenvalues
associated with the onset of localization. The largest $\cT_n$
(smallest $\phi_n$) eigenvalue crystallizes at $\phi=2T$,
corresponding to $\cT\sim 4e^{-2T}=4e^{-2L/\xi}$ if $L\gg \xi$.

\subsection{Mean conductance}

A second, and more direct way of inferring exponential
localization is to investigate the mean (spin) conductance of the
superconducting wire. The mean conductance $\langle \sum \cT_n
\rangle \equiv C(T)$ of the thick disordered wire can be obtained
by integrating $\cT = 1 / \cosh^2 (\phi/2)$ against the eigenvalue
density $\rho_t(\phi) d\phi$ with the result
\begin{equation}
  C(T) = \frac{1}{T} - \frac{1}{3} +
  \sum_{n=1}^\infty \left( \frac{4}{T} +
    \frac{2}{\pi^2 n^2} \right) e^{-n^2 \pi^2 / T} \;.
    \label{result}
\end{equation}
On applying the rescaling $C(T) \to 4 C(4T)$, the latter agrees
with the expression derived in Ref.~\onlinecite{bfgm}. The leading
term $T^{-1}$ is the Ohmic contribution, while the constant $-1/3$
represents the weak localization correction.

To extract the large-$T$ asymptotics and demonstrate exponential
localization, the above expression for $C(T)$ is inconvenient.
Instead, it is convenient to implement an integral transform of
the expression which brings it to a form in which the large-$T$
asymptotics can be developed. Making use of the identities $(1 +
2/x) \, e^{-1/x} = -\int (1/x^2 - 2 / x^3 )\, e^{-1/x} dx $ and
$\sum_{n=1}^\infty n^{-2} = \pi^2 / 6$, one may easily show that
\begin{displaymath}
  C(T) = \sum_{n \in {\mathbb Z}} \int_T^\infty (1 / \tau^2 - 2 \pi^2
  n^2 / \tau^3 ) \, e^{-n^2 \pi^2 / \tau} d\tau \;.
\end{displaymath}
Noting that the function $k \mapsto (1/\tau^2 - k^2 / 2\tau^3) \,
e^{ -k^2 / 4\tau}$ is the Fourier transform of $x\mapsto 2x^2 e^{-
x^2 \tau} / \sqrt{\pi\tau}$, Poisson summation yields
\begin{displaymath}
  C(T) = \frac{4}{\sqrt{\pi}} \sum_{l = 1}^\infty l^2 \int_T^\infty
  e^{-l^2 \tau} \frac{d\tau}{\sqrt{\tau}} \;.
\end{displaymath}
Finally, introducing the complement of the error function,
\begin{displaymath}
  {\rm erfc}(z) = \frac{2}{\sqrt{\pi}} \int_z^\infty e^{-s^2} ds
  = 1 - {\rm erf}(z) \;,
\end{displaymath}
the expression for $C(t)$ may be brought to the final form
\begin{eqnarray}\label{resultalt}
  C(T) &=& \frac{1}{T} - \frac{1}{3} + \sum_{n=1}^\infty \left(
    \frac{4}{T} + \frac{2}{\pi^2 n^2} \right) e^{-n^2 \pi^2 / T}\nonumber\\
  &=& 4 \sum_{l = 1}^\infty l \, {\rm erfc}(l\sqrt{T}) \;.
\end{eqnarray}
While the original expression provides access to the small-$T$
asymptotics, the second one gives easy access to the large-$T$
behaviour, viz.
\begin{displaymath}
  C(T) \stackrel{T\to\infty}{\longrightarrow} \frac{4}{\sqrt{\pi T}}
  \, e^{-T} \;.
\end{displaymath}
Recalling $T= L / \xi$, the latter shows the characteristic
exponential dependence of localization with a localization length
$\xi$.

Consideration of the eigenvalue density and mean conductance
provide strong circumstantial evidence that the expression derived
for the generating function is exact. However, to emphasize the
validity of the semi-classical expansion, it is useful to provide
an explicit calculation of the partition function as a solution of
the Euclidean-time `Schr\"odinger' equation or heat kernel.

\subsection{Heat Kernel}

By canonical quantization, the path integral with the Lagrangian
$\frac{1}{8} \mathrm{STr}\, (g^{-1} \dot g)^2$ becomes the
`Schr\"odinger equation' for free-particle quantum motion on $G$.
The Schr\"odinger operator is the negative of the Laplace-Beltrami
operator $\triangle$ (or Laplacian for short) on $G$. Thus the
heat kernel satisfies a Schr\"odinger-like equation which is the
diffusion (or heat) equation on $G$:
\begin{displaymath}
    \partial_t W(g,g^\prime;t) = \triangle W(g,g^\prime;t)
    \qquad (t > 0) \;,
\end{displaymath}
where $\triangle$ acts on the left argument of $W$. The reduction
from the full heat kernel to the radial function $\omega_t(\phi,\theta)$
on the maximal Abelian subgroup $A$ takes the diffusion equation into
\begin{displaymath}
    \partial_t \omega_t(\phi,\theta) = (\triangle \omega_t)(\phi,\theta)
    \qquad (t > 0)\;;
\end{displaymath}
where we continue to denote the radial part of the Laplacian by
$\triangle$ for simplicity. Since $G$ has superdimension zero
(there being the same number of bosonic and fermionic degrees of
freedom) the short-time asymptotics of the heat kernel is given by
\begin{displaymath}
    \omega_t(\phi,\theta) \stackrel{t \to 0+}{\longrightarrow}
    e^{-(\phi^2 + \theta^2)/4t} \;.
\end{displaymath}
(In dimensions $d \not= 0$, the Gaussian would be preceded by a
factor $(4\pi t)^{-d/2}$.)

To proceed, we must draw on some basic facts from the geometry and
analysis of the supermanifold $G$:
\begin{itemize}

\item Diagonalization of matrices $g$ (i.e.~$g = h a h^{-1}$ with
$a \in A$) defines a polar decomposition of $G$. By this
decomposition, the $G$-invariant Berezin integration measure for
$G$ determines a radial integration measure $J dx dy$, which is
positive on the Weyl chamber chamber $[0,\infty] \times [0,\pi]
\subset {\rm Lie}A$.

\item By a standard formula \cite{helgason} from the theory of Lie
groups and symmetric spaces (for a recent pedagogical review see,
for example, Ref.~\onlinecite{magnea}), the measure function $J$
for the case at hand is given by
\begin{displaymath}
  J = \frac{\sin^2 \theta}{\sinh^2 \left( (\phi+i\theta)/2 \right)
  \sinh^2 \left( (\phi-i\theta)/2 \right)} \;.
\end{displaymath}

\item Finally, the Laplacian on radial functions is given by
\begin{displaymath}
  (\triangle \omega_t)(\phi,\theta) = \left( J^{-1} \partial_\phi J
  \partial_\phi +
  J^{-1} \partial_\theta J \partial_\theta \right) \omega_t(\phi,\theta) \;.
\end{displaymath}

\end{itemize}

By introducing complex coordinates $z = (\phi + i\theta)/2$ and
$\bar z = (\phi - i\theta)/2$ it is apparent that the analytic
square root of the measure function,
\begin{displaymath}
  J^{1/2} = \frac{ \sin(i\bar z - i z) } {\sinh(z)
    \sinh(\bar z)} = i \frac{\cosh(z)}{\sinh(z)} - i
  \frac{\cosh(\bar z)}{\sinh(\bar z)} \;,
\end{displaymath}
is harmonic: $(\partial_\phi^2 + \partial_\theta^2) J^{1/2} = 0$.
Using this property one can easily verify that the radial part of
the Laplacian can be cast in the form
\begin{eqnarray*}
  (\triangle \omega_t)(\phi,\theta) = J^{-1/2} (\partial_\phi \partial_\phi +
  \partial_\theta \partial_\theta ) J^{1/2} \omega_t(\phi,\theta) \;.
\end{eqnarray*}
It therefore follows that the product $E_t = J^{1/2} \omega_t$
satisfies the Euclidean heat equation
\begin{displaymath}
  \partial_t E_t = (\partial_\phi^2 + \partial_\theta^2) E_t \;.
\end{displaymath}

The Euclidean heat kernel in two dimensions is known to be $(4\pi
t)^{-1} e^{- (\phi^2 + \theta^2) / 4t}$. However, this is not the
solution we want here: As mentioned earlier, the heat kernel
$\omega_t(\phi,\theta)$ is subject to zero-dimensional small-$t$
asymptotics, $\omega_t(\phi,\theta) \rightarrow e^{-(\phi^2 +
\theta^2)/4t}$.  This short-time behavior is achieved with the
choice
\begin{displaymath}
  {\tilde E}_t = \frac{4\theta}{\phi^2+\theta^2} \, e^{-(\phi^2+\theta^2)/4t}
  \;.
\end{displaymath}
To confirm that ${\tilde E}_t$ satisfies the Euclidean heat
equation, one uses the identity $\partial_\phi^2 +
\partial_\theta^2 = \partial_z \partial_{\bar z}$ and
\begin{displaymath}\
  {\tilde E}_t = - 2 \, {\rm Im} (z^{-1}) \, e^{-z \bar z / t}\;.
\end{displaymath}
The solution ${\tilde\omega}_t = J^{-1/2} {\tilde E}_t$ thus
obtained is not yet $2\pi$-periodic in $\theta$ and hence does not
lift to a function on the Abelian group $A = {\mathbb R}_+ \times
{\rm U}(1)$. Enforcing periodicity by summing over images, we
obtain:
 \[ \omega_t(\phi,\theta) = J^{-1/2}(\phi,\theta) \sum_{n \in {\mathbb Z}}
  \frac{4 \theta_n}{\phi^2 + \theta_n^2} \, e^{-(\phi^2+\theta_n^2)/4t} \;,\]
where $\theta_n = \theta + 2\pi n$ as before. This is the correct
answer, and it is easily seen to coincide exactly with the
semi-classical result for the partition function derived above.

As emphasized above, the coincidence of the semi-classical
expansion with the exact expression for the partition function is
a particular feature of the group manifold structure of the
$\sigma$-model for the symmetry class $C$I. As a result, we can
deduce the existence of other novel symmetry classes where the
transport properties of the quasi one-dimensional system can be
inferred from the structure of the semi-classical expansion.

\section{Generalizations}

Having established the principle, we now turn to consider
extensions of the present scheme to other symmetry classes. Here
we consider classes $D$III and $A$III:

\subsection{Class $D$III (even)}

Symmetry $D$III is realized \cite{az} in superconductors which
exhibit time-reversal symmetry but where the ${\rm SU}(2)$ spin
rotation symmetry is broken by a spin scattering mechanism such as
a spin-orbit interaction. The latter presents a critical testing
ground for the present theory as it has already been established
by Brouwer et al. \cite{bfgm} that exponential localization is
{\it absent} generically in quasi one-dimensional systems of this
symmetry class.

For the symmetry class $D$III, the corresponding target space of
the nonlinear $\sigma$-model is a Riemannian symmetric superspace
of type $C|D$ \cite{suprev}. This means that we are to work again
with the complex orthosymplectic Lie supergroup, $G = {\rm
OSp}(2n|2n)$, but now one must select a non-compact symmetric
space ${\rm Sp}(2n,\mathbb{C}) / {\rm USp}(2n)$ in the boson-boson
sector and the compact group ${\rm SO}(2n)$ in the fermion-fermion
sector. For $n = 1$ these spaces are isomorphic to ${\rm H}^3$
(the three-hyperboloid) and ${\rm U}(1)$ respectively. A maximal
commuting subgroup $A = {\mathbb R}_+ \times {\rm U}(1)$ is still
formed by diagonal matrices $a = {\rm diag}(e^\phi, e^{-\phi},
e^{i\theta}, e^{-i\theta})$ with $\phi \in{\mathbb R}$ and $\theta
\in [-\pi , \pi]$, so the saddle-point configurations
(\ref{spset}) are unchanged. Only the fluctuation contribution
differs.

The parametrization of the target space by the exponential mapping
$\tilde g=\exp X$ is the same as for class $C$I but with the
boson-boson and fermion-fermion sectors interchanged:
\begin{displaymath}
 X = \begin{pmatrix} e &b &\delta &\beta \\ c &- e &-\gamma &-\alpha
 \\ \alpha &\beta &d &0 \\ \gamma &\delta &0 &-d \end{pmatrix} \;,
\end{displaymath}
where $d$ is now imaginary, $e$ real, and $c = \bar b$. In view of
the duality (by exchange of the compact and non-compact sectors)
connecting the nonlinear $\sigma$-models for the classes $C$I and
$D$III, all calculations for $D$III are very similar to those for
$C$I and, for brevity, we simply quote the results here. The
partition function, obtained by performing the sum over geodesics
together with the integral over Gaussian fluctuations, is given by
\begin{eqnarray}\label{DIIIkern}
  &&\cZ_T(\phi,\theta) = \sum_{n \in {\mathbb Z}} \frac{\sinh \left(
      \frac{1}{2}(\phi + i\theta_n) \right)}{\frac{1}{2} (\phi + {\rm
      i}\theta_n)} \, \frac{\sinh \left( \frac{1}{2}(\phi - i\theta_n)
    \right)}{\frac{1}{2} (\phi - i\theta_n)}\nonumber \\ &&\qquad
    \qquad\qquad\qquad \times\frac{\phi}{\sinh \phi} \,
    e^{-(\phi^2 +\theta_n^2) / 4T} \;.
\end{eqnarray}

{}From Rejaei's relation, the corresponding density of
transmission eigenvalues is then given by
\begin{widetext}
\begin{equation}\label{DIIIdens}
  \rho_T^{D{\rm III}}(\phi) = \frac{1}{2T} + \frac{1}{2(\phi^2 + \pi^2)}
  - \sum_{n \not= 0} \frac{ e^{-n (n+1) \pi^2 /T} }{2\pi^2 n}
  \, {\rm Re} \, \frac{ \phi + i\pi}{ \phi + i\pi (n+1) }
    \, e^{ i n\pi \phi / T} \;.
\end{equation}
\end{widetext}
A plot of this function in Fig.~\ref{fig:densities} shows the
`crystallization' of transmission eigenvalues for $t \gg 1$. Yet,
exponential localization does not take place, as $\rho_T(\phi)$
peaks at $\phi = 0$ (maximal transmission), and the peak amplitude
decays only algebraically with increasing $T$.

Turning to the mean conductance, the partition function yields the
following expression:
\begin{eqnarray}\label{DIIIcond}
  C(T) &=& \frac{1}{T} + \frac{1}{3} - \sum_{n=1}^\infty
  \frac{2}{\pi^2 n^2} \, e^{- n^2 \pi^2 / T} \nonumber \\ &=&
  \frac{2}{\sqrt{\pi T}} + \frac{2}{\sqrt{\pi}} \sum_{l =
    1}^\infty \int_T^\infty e^{-l^2 \tau} \frac{d\tau}
  {\tau^{3/2}} \;.
\end{eqnarray}
The first expression agrees with the result of Ref.~\onlinecite{bfgm}; the second is obtained from it by Poisson resummation. The first two terms, $C^{\rm pert}(T) =t^{-1} + 1/3$,
represent the Ohmic and weak anti-localization terms that can be
computed from a standard perturbation theory for the mean
conductance. As with symmetry class $C$I, higher-order corrections
from the perturbative expansion vanish identically. All non-zero
corrections are non-perturbative and arise from the non-trivial
geodesics in the compact sector. The second expression
in Eq.~(\ref{DIIIcond}) presents an anomalous diffusive
asymptotics for large $T = L / \xi$:
\begin{displaymath}
  C(T) \approx \frac{2}{\sqrt{\pi T}} \qquad (T \gg 1) \;,
\end{displaymath}
a feature already seen in Ref.~\onlinecite{bfgm}.

Finally, expressed in the form of the heat kernel, the solution
for class $D$III is trivially related by the duality discussed
earlier and the validity of the expression for the partition
function above may be confirmed straightforwardly. As a final
application of semi-classical exactness, we turn now to one of the
chiral symmetry classes.

\subsection{Class $A$III}

This symmetry class is relevant to the low-energy physics of the
chiral Dirac operator \cite{verbaar} and to the random flux model
\cite{as}. Lately, it has been discussed \cite{ntw,asz} in the
context of the quasi-particle properties of a $d$-wave
superconductor subject to a smooth random potential. Previous
studies of quasi one-dimensional systems in this
class~\cite{bmsa,mbf,bmf,am} have revealed a surprisingly rich
behaviour, including localization-delocalization transitions with
a number of critical points.

\subsubsection{Rejaei relation}

Before exploring the properties of the nonlinear $\sigma$-model
action, it is first necessary to confirm the form of Rejaei's
relation for this symmetry class. For this we refer to the
$\sigma$-model formulation of the bond disordered quasi
one-dimensional chain given by Altland and Merkt~\cite{am}. We
consider $N$ chains of $2M\gg N$ sites with hopping matrix element
$t^{(m)}_{\mathrm{h}}=1+(-)^m a$ along the chains between sites
$2m$ and $2m+1$, where $a$ is a `staggering' parameter. In the
following, we suppose that a weak disorder potential (which,
further, lifts time-reversal symmetry) of strength  $\ll 1$
couples the chains. We view the single particle Hilbert space as
the product $\mathsf{H} = \mathbb{C}^2 \otimes \mathbb{C}^N
\otimes\mathbb{C}^M$, with the first factor referring to the odd
and even numbered sites along the chain.

The sublattice symmetry of the Hamiltonian translates to the
condition $H=-\Sigma_3H\Sigma_3$, where $\Sigma_i$ represent the
Pauli matrices in this odd/even subspace. From this sublattice
symmetry there follows the relation
\[G^A_{-\epsilon}=-\Sigma_3 G^R_{\epsilon}\Sigma_3\,. \]
The corresponding current operator is simply given by $\Sigma_2/2$
(if we ignore the disorder and staggering, or evaluate it in the
clean, unstaggered leads). Thus the moments of the transmission
matrix at zero energy are given by
\begin{eqnarray}
\mathrm{tr}({\bf t}_0{\bf t}^{\dagger}_0)^n&=&\mathrm{tr}_{\mathsf H}(
\hat v_L G_0^A \hat v_R G_0^R)^n\nonumber\\*
&=&\mathrm{tr}_{\mathsf H}(\cP_1\Sigma_2 G_0^A \cP_{M}\Sigma_2 G_0^R)^n
\nonumber\\*
&=&-\mathrm{tr}_{\mathsf H}(\cP_1\Sigma_1 G_0^R\cP_{M}\Sigma_1 G_0^R)^n,
\end{eqnarray}
where $\cP_m$ projects onto the $2m^{\mathrm{th}}$ site of each
chain. It seems that these moments cannot in general be presented
as the coefficients in some expansion of a functional determinant.
The average conductance ($n=1$) may however be obtained as the
$\phi_L\phi_R$ coefficient of the determinant of a hamiltonian
with hopping $e^{i\phi_L}$ between the $1^{\mathrm{st}}$ and
$2^{\mathrm{nd}}$ sites, and $e^{i\phi_R}$ between the
$2M-1^{\mathrm{th}}$ and $2M^{\mathrm{th}}$ sites (since hopping
is described by a $\Sigma_1$ term in the Hamiltonian).

We may thus study the usual partition function
\begin{displaymath}
  \cZ_T(\phi,\theta) = \int\limits_{g_0={\bf 1}}^{g_T = a(\phi,\theta)
  }{\cal D}g \, \exp \left( -\frac{1}{4} \int\limits_0^T
  {\cal L} dt \right) \;,
\end{displaymath}
for $a(\phi,\theta)=\mathrm{diag}\left( e^{i\phi},e^{i\theta} \right)$ (the
matrix $g$ being $2\times 2$). The conductance can be then identified with
the coefficient $C(T)$ in the expansion
\begin{eqnarray} \label{condrel}
 &&\cZ_T(\phi,\theta) = 1 + C_0(T) (\phi - i\theta) - C(T) (\phi^2 + \theta^2)
  / 4 \nonumber\\
 && \qquad\qquad\qquad\qquad + C_1(T) (\phi - i\theta)^2 + \ldots
\end{eqnarray}

\subsubsection{Semi-classical calculation}

The most general nonlinear $\sigma$-model Lagrangian for quasi
one-dimensional systems belonging to class $A$III (and in suitable
units of length) depends on two parameters $u, v$:
\[  {\cal L} = {\rm STr}\, (g^{-1} \dot g)^2 + 4 v \, {\rm STr} \,
  g^{-1}\dot g - u ({\rm STr} \, g^{-1} \dot g)^2 \;.\]
The target space of the nonlinear $\sigma$-model for this class
\cite{suprev} is of type $A|A$ and is obtained from the complex
Lie supergroup ${\rm GL}(n|n)$ by picking $M_{\rm bb} = {\rm
GL}(n,\mathbb{C}) / {\rm U}(n)$ in the boson-boson sector and
$M_{\rm ff} = {\rm U} (n)$ in the fermion-fermion sector. For $n =
1$ these are ${\rm GL}(1, \mathbb{C}) / {\rm U}(1) \simeq
\mathbb{R}_+$ and ${\rm U}(1)$ respectively.  The exponential
parametrization $\tilde g=\exp X$ of the target space is achieved
by setting
    \[X = \begin{pmatrix} p &\alpha\\ \beta &iq \end{pmatrix}\]
with real commuting variables $p,q$ and anti-commuting variables
$\alpha, \beta$. Again, $a(\phi,\theta)$ parameterizes the maximal
Abelian subgroup. Based on the geodesics $a_t^{(n)}= {\rm diag}
\left( e^{\phi t/T} , e^{i\theta_n t/T} \right)$ with $\theta_n =
\theta + 2\pi n$, the semiclassical approximation for
$\cZ_T(\phi,\theta)$ is found to be
\begin{widetext}
\begin{equation}\label{AIIIkern}
  \cZ_T(\phi,\theta) = \sum_{n \in {\mathbb Z}} (-1)^n
    \frac{ \sinh \big( \frac{1}{2} (\phi - i\theta) \big)}
    {\frac{1}{2}(\phi - i\theta_n)} \,
    e^{- v (\phi - i\theta_n) -[\phi^2 + \theta_n^2 - u ( \phi
    - i\theta_n )^2] / 4T} \;.
\end{equation}
\end{widetext}
Expanding the result above for the partition function, and using
Eq.~(\ref{condrel}) gives
\begin{eqnarray}
  C(T) &=& \frac{1}{T} + \frac{2}{T} \sum_{n = 1}^\infty (-1)^n
  \cos(2\pi n v) \, e^{- (1 + u) n^2 \pi^2 / T} \nonumber \\
  &=& \frac{1}{\sqrt{ \pi T (1 + u)}} \sum_{l \in {\mathbb Z}}
  e^{- (l + v - 1/2)^2 T / (1 + u)} \;, \label{AIIIcond}
\end{eqnarray}
where the second expression is obtained by Poisson resummation of the
first.

The result (\ref{AIIIcond}) depends periodically on the parameter
$\theta \equiv 2\pi v$ with period $2\pi$. In fact, $\theta$ has
the meaning of a topological angle. From Ref.~\onlinecite{am} we
know that tuning $\theta$ from $0$ to $2\pi$ amounts to changing
by two the number of channels (which is assumed to be large in
order for the $\sigma$-model approximation be valid).  In terms of
the bond disordered chains discussed earlier
\[\theta=\pi\left( N-f \right) \quad \mathrm{mod}\, 2\pi,\]
where $f$ is the staggering parameter $a$ scaled by the strength
of the disorder. We see that without staggering, a delocalized
state appears only for odd channel number. The localization length
diverges with critical exponent $1$ as the staggering approaches
the critical values $f=N-1\,\left( \mathrm{mod} \,2 \right)$.

If class $A$III is realized by Dirac fermions in a random Abelian
gauge field, the parameter $u > 0$ measures the strength of the
gauge disorder. Although the Lagrangian ${\cal L}$ at first sight
would seem to become unstable at $u = 1$ (the coefficient of $\dot
p^2$ becomes negative there), this is not really so. What matters
is the \emph{full} quadratic form
\begin{displaymath}
    {\cal L}_{v = 0} = (1 - u) \dot p^2 + (1 + u) \dot q^2 +
    2i u \dot p \dot q + \mbox{odd variables} \;,
\end{displaymath}
which has the Jordan normal matrix form $\begin{pmatrix} 1 &u\\ 0
&1 \end{pmatrix}$, and is invertible for any $u$. In fact, the
final result (\ref{AIIIcond}) reveals a smooth dependence of the
mean conductance on $u$ for all physical values ($u > 0$)
including the fake singularity $u = 1$.

\subsubsection{Heat Kernel}

Finally, seeking an exact solution through the heat kernel, an
exact calculation of $\omega_T(\phi,\theta)$ is still possible as
the Hamiltonian ${\cal H}$ obtained by canonical quantization from
${\cal L}$ can once again be transformed to Euclidean form:
\begin{eqnarray*}
  &&{\cal H} = - J^{-1/2} \left( \partial_\phi \partial_\phi + \partial_\theta
    \partial_\theta + 2v (\partial_\phi - i\partial_\theta) \right.
    \nonumber \\ &&\qquad\qquad\qquad\qquad\qquad \left.+
    u (\partial_\phi - i\partial_\theta)^2 \right) J^{1/2} \;,
\end{eqnarray*}
with $J = 1 / \sinh^2 \left( \frac{1}{2} (\phi - i\theta) \right)$.
By solving the differential equation $\partial_t \omega_t = -
{\cal H} \omega_t$ with the appropriate $\delta$-function initial
conditions at $t = 0+$, we exactly recover the semiclassical
answer (\ref{AIIIkern}).

\section{Conclusion}

We have confirmed the exactness of our semiclassical analysis for
symmetry classes $C$I, $D$III, and $A$III. By simply summing over
the saddle points of the classical action --- recall that these
differed by the length of the geodesic looping around the compact
fermion-fermion submanifold of the theory --- and treating the
fluctuations in the Gaussian approximation, we have obtained the
exact result for the partition function.

The underlying reason for the exactness of our calculations is that the path integral on the $\sigma$-model manifolds considered here satisfies an infinite-dimensional generalisation of the Duistermaat-Heckman theorem~\cite{picken,szabo}. That is, the integration manifold in Eq.~\ref{g_action}, being related to the space $\Omega G$ of based loops in $G$, is symplectic with Liouville measure, and the action is the momentum mapping on this manifold. 

We have discussed all the symmetry classes where the
$\sigma$-model target space has a group-like structure, and thus
exhausted the situations in which the semiclassical approach is
exact. Nevertheless, we hope that this approach may inform future
investigations of more taxing localization problems. One very
interesting extension that could be tackled in a similar manner is
the critical scaling of the localization length near the band
center in situations where zero energy is described by class
$D$III or $A$III. This lies outside of the scope of methods like
the DMPK equation.

\appendix*
\section{Embedding of $g$ in $Q$ for Class $C$I} \label{app:embedding}

Let us review the structure of the $\sigma$-model target space for
Class $C$I. The construction used in Ref.~\onlinecite{suprev} is
based on `copying' the symmetries of the Hamiltonian to symmetries
of the auxiliary space of the $Q$-field. For Class $C$I
(particle-hole and time reversal symmetry), this is achieved by
demanding that the fields satisfy
\begin{eqnarray} \label{matrixcond}
        &&\Psi_M = {\cal C} \bar\Psi_M^{\rm T} \gamma^{-1}, \quad
        \bar\Psi_M = - \gamma \Psi_M^{\rm T} {\cal C}^{-1}, 
	\nonumber \\
        &&\Psi_M = \bar\Psi_M^{\rm T} \tau^{-1}, \quad
        \bar\Psi_M = \tau \Psi^{\rm T} .
\end{eqnarray}
The subscript $M$ is to remind us that in the formulation of
Ref.~\onlinecite{suprev} the variables $\Psi_M$ and $\bar\Psi_M$
are (super)\emph{matrices} mapping the auxiliary space to the
physical Hilbert space (where the Hamiltonian acts), and vice
versa. In the present case we work with the more traditional
column and row supervectors $\Psi_V$ and $\bar\Psi_V$, since:

\begin{itemize}
\item In the localization problem the $Q(\mathbf{r})$ field used to
decouple the term arising from potential disorder retains the
Hilbert space structure of the Hamiltonian (particle hole space
and spatial index), whereas in the random matrix problem that
structure is lost on averaging and, more importantly,
\item We use some of the additional structure (the cc space) to enlarge
our Hamiltonian (Eq.~\ref{enlargedH}).
\end{itemize}
Transcribing the conditions (\ref{matrixcond}) for supervectors gives
\begin{eqnarray} \label{vectorcond}
        &&\Psi_V =\gamma {\cal C} \bar\Psi_V^{\rm T}, \quad
        \bar\Psi_V = - \Psi_V^{\rm T} {\cal C}^{-1}\gamma^{-1},
        \nonumber \\
        &&\Psi_V = \tau\bar\Psi_V^{\rm T} , \quad
        \bar\Psi_V =  \Psi^{\rm T}\tau^{-1} ,
\end{eqnarray}
where we used the fact that $\gamma$ and $\tau$ can be chosen to
be orthogonal matrices. Now in the localization problem
$Q(\mathbf{r})$ is a $16\times 16$ matrix
(ph$\times$cc$\times$tr$\times$bf - by `tr' we mean the space to
accommodate time reversal symmetry), so we need to check that the
conditions (\ref{vectorcond}) give rise to a $\sigma$-model with
the same $8\times 8$ space as in the random matrix problem. This
is straightforwardly verified following
Ref.~\onlinecite{zirnsuper} by observing that $Q(\mathbf{r})\sim
\Psi\otimes\bar\Psi\sigma_3$ (we drop the $V$ subscript from now
on), arising from a decoupling of potential disorder, has the
symmetry
\begin{eqnarray} \label{Qsym}
&&Q = \sigma_1\gamma Q^T\gamma^{-1}\sigma_1 \;,
\nonumber\\
&&Q = \sigma_3\tau Q^T\tau^{-1}\sigma_3 \;.
\end{eqnarray}
The saddle-point manifold is spanned by
\[Q=W\sigma_3\Sigma_3W^{-1},\qquad W=w\otimes\openone_{\mathrm{ph}},\]
or
\[ Q=\sigma_3 q,\qquad q=w\Sigma_3 w^{-1}.\]
Thus (\ref{Qsym}) implies that the symmetries of the $8\times 8$
field $q$ are
\begin{eqnarray} \label{qsym}
q=-\gamma q^T\gamma^{-1}, \nonumber\\*
q=\tau q^T\tau^{-1},
\end{eqnarray}
which are the same relations found in Ref.~\onlinecite{suprev}. To
construct an embedding of the $4\times 4$ matrix group-valued
field $g$ in $q$ requires an explicit choice of $\gamma$ and
$\tau$. We use
\begin{eqnarray*}
    \gamma &=& E_{\mathrm{bb}} \otimes \gamma_{\mathrm{b}} +
    E_{\mathrm{ff}} \otimes \gamma_{\mathrm{F}}, \\
    \tau &=& E_{\mathrm{bb}} \otimes \tau_{\mathrm{b}} +
    E_{\mathrm{ff}} \otimes \tau_{\mathrm{f}},
\end{eqnarray*}
\begin{eqnarray*}
    \gamma_{\mathrm{b}} = \Sigma_1 \otimes \tau_3 ,
    \quad \gamma_{\mathrm{f}} = i\Sigma_2 \otimes
    \openone_{\mathrm{tr}} , \nonumber \\*
    \tau_{\mathrm{b}} = \openone_{\mathrm{cc}} \otimes \tau_1  ,
    \quad \tau_{\mathrm{f}} = \Sigma_3 \otimes i\tau_2  ,
\end{eqnarray*}
where $\tau_i$ are the Pauli matrices in tr space, and
$E_{\mathrm{bb}}$ and $E_{\mathrm{ff}}$ are projectors in the
bose-bose and fermi-fermi sectors. Note that the structure of the
vector in cc space in the bosonic sector is in fact then
\emph{not} given by Eq.~\ref{ccdoubling}, but rather looks like
\begin{equation*}
\Psi_B=\frac{1}{\sqrt{2}}\begin{pmatrix} \psi_B \cr \cC\tau_3\bar\psi_B^T
\end{pmatrix}_{\mathrm{cc}}, \qquad \bar\Psi_B=\frac{1}{\sqrt{2}}
\left(\bar\psi_B, -\psi_B^T\tau_3\cC^{-1}\right)_{\mathrm{cc}}.
\end{equation*}
Fortunately this doesn't change the boundary conditions. Now the
correct embedding is obtained by finding a transformation to
diagonalize the matrix
\[\eta\equiv-i\gamma\tau^{-1}=\Sigma_1\otimes\tau_2
\otimes\openone_{\mathrm{bf}} , \]
\[\eta\longrightarrow U^\dagger\eta U=\begin{pmatrix}
1 & 0 & 0 & 0 \cr
0 & 1 & 0 & 0 \cr
0 & 0 & -1 & 0 \cr
0& 0& 0& -1
\end{pmatrix}\otimes\openone_{\mathrm{bf}},\]
while simultaneously sending
 \[\Sigma_3\otimes\openone_{\mathrm{tr}}\otimes
 \openone_{\mathrm{bf}}\longrightarrow
 \begin{pmatrix}
0 & 0 & -1 & 0 \cr
0 & 0 & 0 & -1 \cr
-1 & 0 & 0 & 0 \cr
0& -1& 0& 0
\end{pmatrix}\otimes\openone_{\mathrm{bf}}.
 \]
Why? The point is that the set of $w$'s that preserve the
symmetries (\ref{qsym}) takes the form
\begin{equation} \label{tdef}
w\longrightarrow\mathrm{diag}(g_1,g_2),
\qquad g_1,g_2\in \mathrm{OSp}(2|2),
\end{equation}
in such a basis. The stability group of transformations that send
$w\Sigma_3 w^{-1}\to \Sigma_3$ is then given by Eq.~\ref{tdef}
with $g_1=g_2$, so that the saddle-point manifold is parameterized
by
\[q=U\mathrm{diag}(g,g^{-1})U^\dagger\Sigma_3 ,
\qquad g\in \mathrm{OSp}(2|2)\;.\]
An explicit form for $U$ is
\[U=\frac{1}{2}\begin{pmatrix}
-i & -1 & i & 1 \cr
1 & i & -1 & -i \cr
-i & 1 & -i & 1 \cr
1& -i& 1& -i
\end{pmatrix}\otimes\openone_{\mathrm{bf}}.\]
Let us verify that this takes the maximal abelian subgroup $A$ to
the $q$ matrices of the form (\ref{bc}):
\begin{eqnarray*}
q&=&U\mathrm{diag}(a(\phi,\theta),a^{-1}(\phi,\theta))U^{\dagger}\Sigma_3 \\*
&=&\mathrm{diag}\left(\begin{pmatrix} \cosh \phi  & -\sinh \phi \cr \sinh \phi
&-\cosh \phi \end{pmatrix},\begin{pmatrix} \cos \theta & -i\sin \theta \cr i\sin
\theta & -\cos \theta \end{pmatrix}\right)_{\mathrm{bf}}\otimes\openone_{\mathrm{tr}}.
\end{eqnarray*}
Finally, the form of the action (\ref{g_action}) in terms of $g$ follows from
\begin{eqnarray*}
\mathrm{STr}\left[ \nabla q \right]^2&=&2\,\mathrm{STr}\left[ \nabla g\nabla g^{-1} \right]\\
&=&-2\,\mathrm{STr}\left[ g^{-1}\nabla g \right]^2.
\end{eqnarray*}

\end{document}